\documentclass[a4paper]{article}

\usepackage{xcolor}
\usepackage{float}
\usepackage{lineno}
\usepackage[draft]{changes}

\definecolor{ceruleanblue}{rgb}{0.16, 0.32, 0.75}

\definecolor{wongblue}{rgb}{0.0,0.44705883,0.69803923}
\definecolor{wongyellow}{rgb}{0.9019608,0.62352943,0.0}
\definecolor{wonggreen}{rgb}{0.0,0.61960787,0.4509804}
\definecolor{wongpink}{rgb}{0.8,0.4745098,0.654902}
\definecolor{wonglightblue}{rgb}{0.3372549,0.7058824,0.9137255}
\definecolor{wongorange}{rgb}{0.8352941,0.36862746,0.0}
\definecolor{wonglightyellow}{rgb}{0.9411765,0.89411765,0.25882354}

\colorlet{C0}{wongblue}
\colorlet{C1}{wongyellow}
\colorlet{C2}{wonggreen}
\colorlet{C3}{wongpink}
\colorlet{C4}{wonglightblue}
\colorlet{C5}{wongorange}
\colorlet{C6}{wonglightyellow}

\usepackage[colorlinks=true,
			allcolors=ceruleanblue]{hyperref}

\usepackage[
    backend=biber, 
    style=phys, 
    eprint=true,
    sorting=none,  
    biblabel=brackets,
    bibencoding=utf8]{biblatex}
\usepackage[margin=2cm]{geometry}

\usepackage{mathtools}

\usepackage{subcaption}
\usepackage{amsmath}
\usepackage[concrete]{fontsetup}
\usepackage{xeCJK}
\usepackage{pgfplots}
\pgfplotsset{compat=newest}
\usepackage{tikzscale}
\usetikzlibrary{external,calc,decorations.pathmorphing,patterns}
\makeatletter \newcommand{\pgfplotsdrawaxis}{\pgfplots@draw@axis} \makeatother
\pgfplotsset{axis lines on top/.style={
  axis line style=transparent,
  ticklabel style=transparent, tick style=transparent,
  axis on top=false,
  after end axis/.append code={
    \pgfplotsset{axis line style=opaque, ticklabel style=opaque,
                 tick style=opaque, grid=none}
    \pgfplotsdrawaxis}
}}
\pgfplotsset{every axis/.append style={
        xtick align=outside,
	    ytick align=outside,
	    xtick pos=left,
	    ytick pos=left,
	    major tick length=1mm,
	    major tick style={line width=0.75pt,black},
	    axis line style={line width=0.75pt},
	    minor tick length=0.75mm,
        minor tick style={line width=0.5pt},
        axis on top
        }
    }

\newcommand{\keywords}[1]{%
  \par\vspace{0.5em}
  \noindent{\small\textbf{Keywords:} #1}
}

\usepackage{graphicx} 

\usepackage{textcomp}
\usepackage{physics}
\usepackage{parskip}
\usepackage{appendix}
\usepackage{orcidlink}
\usepackage{authblk}

\def\la{\langle}
\def\ra{\rangle}

\title{Negative Differential Heat Conductivity in a Harmonic Chain Coupled to a Particle Reservoir}

\author{Simon Krekels \orcidlink{0000-0003-1726-6625}}
\author{Christian Maes \orcidlink{0000-0002-0188-697X}} 
\author{Ion Santra \orcidlink{0000-0002-9772-2880}}
\author{Ruoxun Zhai (翟若迅) \orcidlink{0009-0001-7591-3379}}

\affil{Department of Physics and Astronomy, KU Leuven}

\begin{document}

\maketitle


\begin{abstract}
When coupling thermal baths at different temperatures, negative differential thermal conductivity is typically attributed to nonlinear interactions in the connecting medium. In this work, we demonstrate that such an effect can arise purely from the nature of the thermal baths and their coupling with the medium.
Specifically, we construct a bath composed of overdamped thermal particles, which is coupled to one end of a harmonic chain, while the other end is connected to a standard Langevin heat bath. By analyzing the steady-state heat current, we observe significant negative differential thermal conductivity. In particular, as the temperature difference between the two baths diverges, the steady-state heat current through the chain vanishes.
The effect is thermokinetic: we compute the effective dissipative coefficient and we find that it scales inversely with the square of the temperature of the particle bath in the high-temperature limit, resulting in an asymptotic decoupling between the bath and the chain. Our results highlight that nonequilibrium transport properties can be strongly influenced by the structure of the environment and its coupling to the system, even in otherwise linear systems.
\end{abstract}


\section{Introduction}

Transport of energy is a central theme in both technological applications and nonequilibrium statistical mechanics \cite{dhar2001heat,lepri2016thermal,deGrootMazur}. While macroscopic laws of heat conduction are well established, their microscopic origin remains subtle and often depends sensitively on how a system is coupled to its environment. In particular, the nature of the thermal reservoirs and the mechanism by which they exchange energy with a conducting medium can strongly influence transport properties. This becomes especially relevant when different types of degrees of freedom such as particles and elastic modes are coupled through the same medium, as is common in soft and biological matter, for instance, in the interaction between particles and a viscoelastic structure like the cell cortex~\cite{valberg1987magnetic,caspi2000enhanced,le2001motor}.

\textcolor{black}{In this paper, we introduce a minimal microscopic model of a `particle bath' that makes the coupling between particle and elastic degrees of freedom explicit, and use it to study heat transport to a standard Langevin heat bath through a harmonic chain, serving as an idealized elastic string.} 
The particle bath interacts locally with the chain, effectively ``bombarding'' one end of the medium. This setup is inspired by soft-matter systems in which particles are embedded in or coupled to deformable substrates (similar interactions were used to couple particles and fields, see eg.,~\cite{demery2023non,beyen2025coupling}). The interaction is modeled in the spirit of Bernoulli--Euler elasticity, where the chain has a bending energy and the particles exert forces that are linear in the local displacement, akin to a Hookean coupling but sensitive to local curvature.

\textcolor{black}{We find that this seemingly simple modification of the system-bath coupling qualitatively changes the transport behavior. Even though the chain itself is linear, the system exhibits a far-from-equilibrium regime of negative differential heat conductivity: increasing the temperature of the particle reservoir can lead to a decrease of the stationary heat current. We identify the cause of this effect by finding the effective dynamics induced by the particle bath. We eliminate the degrees of freedom of the particle bath, and show that they generate a temperature-dependent friction at the boundary oscillator which grows in such a way as to suppress energy transport at higher temperatures. We refer to this mechanism as a thermokinetic effect, emphasizing its kinetic (rather than structural) origin.  Such negative differential conductance has been seen for various systems, including particle \cite{leitmann2013nonlinear,baerts2013frenetic}, and energy transport \cite{Zhang_2025,PhysRevB.80.104302,li2006negative}, but the basis of the present study is the well-known and simple paradigmatic harmonic chain with linear Hamiltonian dynamics in the bulk \cite{rieder1967properties}. }

\textcolor{black}{The paper is organized as follows. Section~\ref{sec:setup} introduces the particle bath model. Section~\ref{sec:single_probe} derives the reduced dynamics of the particle coupled to the particle bath, and Sec.~\ref{sec:chain} analyzes the non-monotonic heat current through the full system. The main findings are summarized in the conclusion.}


\begin{figure}
\centering
    \tikzsetnextfilename{drawing}%
    \begin{tikzpicture}
    \coordinate (cBP) at (-0.5,0);
    \coordinate (cq1) at (2,0.5);
    \coordinate (cq2) at (4, -0.5);
    \coordinate (cq3) at (6, 0.25);
    \coordinate (cq4) at (8, -0.25);
    \coordinate (cT) at (10.5, 0);

    \coordinate (p0) at ($(cBP.center) + (-0.75cm, 0)$);
    \coordinate (p1) at ($(cBP.center) + (-0.5cm, 0.1cm)$);
    \coordinate (p2) at ($(cBP.center) + (-0.25cm, -0.1cm)$);
    \coordinate (p3) at ($(cBP.center) + (0.0cm, 0.2cm)$);
    \coordinate (p4) at ($(cBP.center) + (0.25cm, -0.2cm)$);
    \coordinate (p5) at ($(cBP.center) + (0.5cm, 0.075)$);

    \node[rectangle, rounded corners=2.5mm, inner sep=1.cm, thick,draw=C0, fill=C0!20] (BP) at (cBP) {};
    \node at ($(cBP.center) + (135:0.75cm)$) {$x$};
    \node[anchor=south] at (BP.south) {$q_1,\,T_1$};

    \node[rectangle, rounded corners=2.5mm, inner sep=1.cm, thick,draw=C5, fill=C5!20] (T) at (cT) {};
    \node at (T.center) {$T_2$};
    \node[anchor=south west] at (T.south west) {$q_N$};

    \draw[gray, densely dashed] (BP.east) -- (T.west);

    \fill[pattern=north east lines, pattern color=gray!50] (-1.75, -1.5) rectangle (11.75, -1.8);
    \draw[thick, rounded corners=1pt] (-1.75, -1.8) -- (-1.75, -1.5) -- (11.75, -1.5) -- (11.75,-1.8);

    \draw[decoration={aspect=0.3, segment length=2*0.75mm/1.5mm, amplitude=2mm,coil},decorate,gray]
        (-0.5, -1.5) -- (BP.south);
    \draw[decoration={aspect=0.3, segment length=2*2.0mm/1.5mm, amplitude=2mm,coil},decorate,gray]
        (2, -1.5) -- ($(cq1)-(0,0.05)$);
    \draw[decoration={aspect=0.3, segment length=2*1.0mm/1.5mm, amplitude=2mm,coil},decorate,gray]
        (4, -1.5) -- ($(cq2)-(0,0.05)$);
    \draw[decoration={aspect=0.3, segment length=2*1.75mm/1.5mm, amplitude=2mm,coil},decorate,gray]
        (6, -1.5) -- ($(cq3)-(0,0.05)$);
    \draw[decoration={aspect=0.3, segment length=2*1.25mm/1.5mm, amplitude=2mm,coil},decorate,gray]
        (8, -1.5) -- ($(cq4)-(0,0.05)$);
    \draw[decoration={aspect=0.3, segment length=2*0.75mm/1.5mm, amplitude=2mm,coil},decorate,gray]
        (10.5, -1.5) -- (T.south);

    \draw[line width=0.5mm, C1, line cap=round] plot[smooth, tension=0.5] coordinates {
        (p0)
        (p1)
        (p2)
        (p3)
        (p4)
        (p5)
        (BP.east)
        (cq1) (cq2) (cq3) (cq4)
        (T.west)
        ($(T.west)+(0.5,0)$)
    };
    \fill[C0] ($(p1) - (0,1.25mm)$ ) circle (0.75mm);
    \fill[C0] ($(p2) + (0,1.25mm)$ ) circle (0.75mm);
    \fill[C0] ($(p3) - (0,1.25mm)$ ) circle (0.75mm);
    \fill[C0] ($(p4) + (0,1.25mm)$ ) circle (0.75mm);
    \fill[C0] ($(p5) - (0,1.25mm)$ ) circle (0.75mm);

    \fill[fill=C1] (cq1) circle (0.5mm);
    \fill[fill=C1] (cq2) circle (0.5mm);
    \fill[fill=C1] (cq3) circle (0.5mm);
    \fill[fill=C1] (cq4) circle (0.5mm);
    \fill[fill=C1] ($(T.west)+(0.5,0)$) circle (0.5mm);
    \draw[|<->] ($ (cq1) + (-3mm, 0) $) -- ($ (cq1) + (-3mm, -0.5) $) node[midway, left] {\small$q_{2}$};
    \draw[<->] ($ (cq2) + (0, 0) $) -- ($ (cq2) + (0, 0.5) $) node[midway, left] {\small$q_{3}$};
    \draw[|<->] ($ (cq3) + (3mm, 0) $) -- ($ (cq3) + (3mm, -0.25) $) node[below right] {\small$q_{4}$};
    \draw[<->] ($ (cq4) + (0, 0) $) -- ($ (cq4) + (0, 0.25) $) node[above left] {\small$q_{5}$};
\end{tikzpicture}%

    \caption{Schematic of the setup with the particle bath on the left, and the Langevin bath on the right. The membrane is modeled by a succession of springs with displacement $q_{1\ldots N}$.}
    \label{fig:chaf}
    
\end{figure}
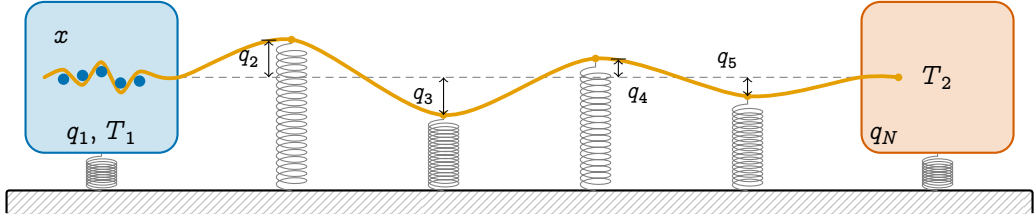

\section{Setup}\label{sec:setup}
We consider heat transport through an elastic string, modeled as a one-dimensional chain of $N$ coupled harmonic oscillators with positions $\{q_j\}$ and momenta $\{p_j\}$, $j=1,\dots,N$. 
This fits within the more general study of heat conduction networks \cite{maes2006heat}, which can be treated in a similar way.

The left boundary oscillator is coupled to a reservoir of overdamped particles with positions $\{x_i\}$ (which we refer to as the particle bath from here on), while the right boundary oscillator is connected to a Langevin heat bath. The Hamiltonian of the oscillator chain, for given bath configuration $\{x_i\}$, is
\begin{equation}\label{eq:chain_hamiltonian}
    \mathcal H= \sum_{j=1}^{N}\frac{p_j^2}{2m} +\sum_{j=1}^{N-1} \frac{k}{2}(q_{j+1}-q_j)^2 + \frac{K}{2} (q_1^2 + q_N^2)
+    \varepsilon q_1\sum_{i=1}^{N_b}\mathcal F(x_i)-\Lambda q_1
\end{equation}
where $\Lambda =\varepsilon\lambda N_b$ characterizes a compensating force, introduced to cancel the constant part of the mean force exerted by the particle bath (cf. Fig.~\ref{fig:gamma_eff-QS}a), and will be properly introduced in the next section.
This compensation has no effect on the current through the string.
The particle bath consists of $N_b$ overdamped Brownian particles at temperature $T_L$ moving on a ring of length $L$, 
\begin{align}\label{bpa}
    \zeta\dot{x}_i &=-\varepsilon q_1\,\mathcal F'(x_i) + \sqrt{2\zeta T_L}\,\xi_i, \qquad i=1,\ldots,N_b
\end{align}
where $\xi_i=\xi_i(t)$ are standard Gaussian white noise. 
Each bath particle interacts locally with the oscillator $q_1$ through a periodic coupling $F(x)$, which we choose to be the von Mises distribution,
\begin{align}
    \mathcal F(x) = \frac{\lambda}{I_0(p)} \exp\left[p \cos\,\left(\frac{2\pi x}{L}\right)\right],
\end{align}
where $I_0$ is the zeroth modified Bessel function of the first kind. The parameter $p$ characterizes the spatial localization of the interaction, and reduces to a periodic Dirac-delta form in the limit $p\to\infty$.

The equations of motion for the oscillator chain follow,
\begin{equation}\label{eq:EOMq1}
\begin{aligned}
     m\ddot{q}_1 &= -K q_1-k(q_1-q_{2})+\varepsilon \sum_{i=1}^{N_b} \mathcal F(x_i)+N_b\Lambda\\
      m\ddot{q}_i &= k( q_{i-1}-2q_i+q_{i+1})\\
       m\ddot{q}_N &=-\gamma_{R}\dot{q}_N-K q_N-k( q_N-q_{N-1})+\sqrt{2\gamma_RT_R}\,\eta_R
\end{aligned}
\end{equation}
where $\gamma_R$ and $\eta_R$ are the friction and Gaussian white noise respectively, coming from the Langevin bath on the right at temperature $T_R$.

We are interested in the nonequilibrium steady state (NESS), particularly in the heat current flowing through the chain \added{\cite{rieder1967properties}}
\begin{equation}
J = \frac{1}{2m}\la\mathcal J(t)\ra=\left\la (p_j(t)+p_{j+1}(t))(q_{j+1}-q_j)\right\ra
\label{eq:current}
\end{equation}
where the averages are taken in the NESS obtained from solving the full stochastic dynamics \eqref{bpa}--\eqref{eq:EOMq1}.

For harmonic chains coupled to Langevin heat baths at both ends, the heat conduction problem was solved exactly in the seminal work of Rieder, Lieb, and Lebowitz~\cite{rieder1967properties}. 
In that case, the stationary current was found to be independent of the system size, and  proportional to the temperature difference at the two ends. 
In contrast, the particle reservoir in our setup is not a standard thermal bath modeled by independent oscillators or oscillator chains. 
In fact, the interaction is not of the usual two-body form depending on relative displacements, but rather represents a local forcing exerted by particles that effectively bombard the left-most boundary of the chain. 
The induced friction and noise by such a bath have never been addressed in the literature, and is expected to have different properties than oscillator baths. 
To characterize these, we first analyze the reduced problem of a single probe coupled to the bath of overdamped Brownian particles as discussed above. 
This allows us to estimate the effective force, friction, and noise induced by the bath, which we then use to understand transport in the full chain.

\section{Induced forces on a single probe}\label{sec:single_probe}
\begin{figure}
    \centering
    \tikzsetnextfilename{gamma_eff-F_QS}%
    \input{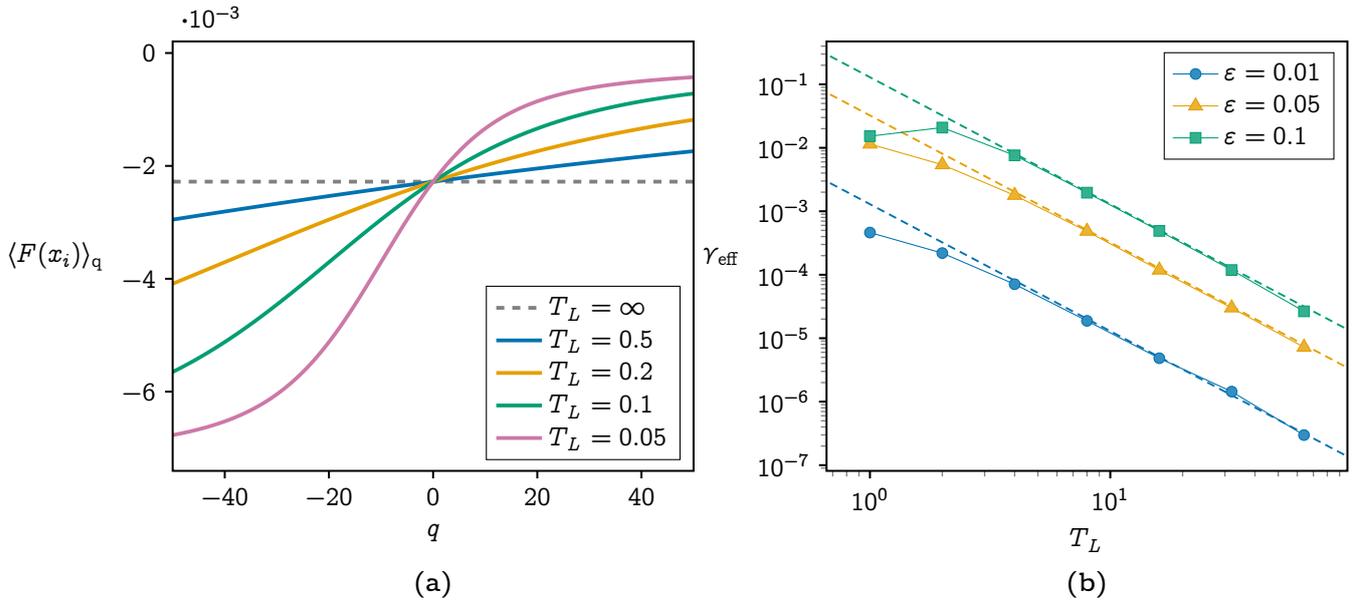}%

    \caption{\textbf{(a)} Mean force exerted by the particle bath at fixed $q$ and temperature $T_L$, with $\varepsilon=10^{-3}$. \textbf{(b)} Plot of the effective friction experienced by a single oscillator in the embedded bath with the bath temperature. The points denote the $\gamma_\text{eff}$ extracted from the simulations using Eq.~\ref{eq:veldecay}, while the dashed lines denote the prediction  Eq.~\ref{eq:gammaGK} for $m=1$, $k=1$, $N_b=40$, $\zeta=0.1$, $\lambda=1$, $p=2$, and $L=25$.}
    \label{fig:gamma_eff-QS}
\end{figure}
In this section, we consider a single oscillator coupled to the bath, corresponding to setting $k=0$ in the previous section.
The dynamics of that single oscillator is then governed by
\begin{align}
m \ddot q &= -K q - \varepsilon \sum_{i=1}^{N_b} F(x_i), \\
\zeta \dot x_i &= -\varepsilon q F'(x_i)
+ \sqrt{2\zeta T_L}\,\xi_i.
\end{align}
We are interested in the regime of time-scale separation, where the bath particles relax much faster than the oscillator.
In this limit, the particle degrees of freedom can be integrated out to obtain an autonomous equation for the slow probe dynamics $q(t)$.
It has the form,
\begin{equation}
m \ddot q(t) = -K q(t) + \sum_{i=1}^{N_b} \left(F^{\text{qs}}_i(t)+F^{\text{fric}}_i(t)+\eta_i(t)\right).
\label{eq:gen_red}
\end{equation}
The terms in the bracket can be identified as the quasistatic mean force, the frictional force, and the fluctuating force, respectively. They are given by, 
\begin{align}
    F_i^{\mathrm{qs}}(t)&=-\varepsilon\la \mathcal F(x_i)\ra_{q(t)},\\
    F_i^{\mathrm{fric}}(t)&=-\varepsilon\Big(
\la \mathcal F(x_i(t)) | q(s), s\le t\ra
-
\la\mathcal  F(x_i)\ra_{q(t)}
\Big)\\
\eta_i(t)&=-\varepsilon\Big(
\mathcal F(x_i(t))
-
\la \mathcal F(x_i(t)) | q(s), s\le t\ra
\Big).
\end{align}
where $\la\cdots\ra_{q(t)}$ denotes averages in the ensemble where $q$ is held at a fixed value.
$F_i^{\mathrm{qs}}(q)$ is displayed in Fig.~\ref{fig:gamma_eff-QS}a.
The above correlations are in general difficult to compute, but yield closed forms in the limit of weak coupling $\varepsilon\to 0$ [see Appendix~\ref{fric} for details].
The mean force is given as,
\begin{equation}
F_i^{\text{qs}}(q)
=-\varepsilon \lambda + \varepsilon^2\lambda^2\,\frac{ q(t)}{T_L }\,
\left(\frac{I_0(2p)}{I_0(p)^2}-1\right)
+O(\varepsilon^3),
\label{eq:fqs_weak}
\end{equation}
where the first term is a constant mean force which cancels the external drive $N_b\Lambda$.
The second term, proportional to $q(t)$, offers a correction to the probe stiffness.
The second term reduces to $F_i^{\text{fric}}(t) = -\gamma_i^{\mathrm{eff}}\,\dot q(t),$ with the effective friction coefficient 
\begin{equation}
\gamma_i^{\text{eff}} = \frac{\varepsilon^2}{T_L} \int_0^{\mathrlap{\infty}} \dd{t} \langle \mathcal F(x(t));\mathcal F(x(0))\rangle_0=\varepsilon^2\frac{\zeta L^2}{2\pi^2T_L^2}\left(\frac{\lambda}{I_0(p)}\right)^2\sum_{n=1}^\infty\frac{I_n(p)^2}{n^2}
\label{eq:gammaGK}
\end{equation}
where we used the force-force correlation
\begin{align}
    \la \mathcal F(x(t));\mathcal F(x(s))\ra_0=\frac{2\lambda^{2}}{I_0(p)^2}\sum_{n=1}^{\infty}I_n(p)^2e^{-\frac{4\pi^2n^2T_L}{\zeta L^2}|t-s|}.
\end{align}
The fluctuating force also satisfies the fluctuation-dissipation relation  $\la\eta_i(t)\eta_j(s)\ra=2\gamma^i_\text{eff}T_L\delta_{ij}\delta(t-s)$. 

The effective friction can also be extracted numerically from the relaxation of the single probe.
The reduced dynamics is that of an underdamped harmonic oscillator, for which the mean velocity decays as
\begin{equation}
\la \dot q(t)\ra=\la \dot q(0)\ra
e^{-\gamma_{\text{eff}}t/(2m)}
\times
\left[\text{oscillatory terms}\right].
\label{eq:veldecay}
\end{equation}
The renormalized stiffness plays a role in governing the oscillatory terms, but does not affect the decay envelope.
We therefore estimate $\gamma_{\text{eff}}$ from the exponential envelope of the numerically measured velocity relaxation starting from $v(t=0)=50$.
Figure~\ref{fig:gamma_eff-QS}b shows a comparison of the predicted friction and that obtained from numerical simulations of the full system, and shows a good agreement for large $T_L$.

\section{Oscillator chain}\label{sec:chain}
\begin{figure}
    \centering
    \tikzsetnextfilename{nondim_comparison}%
    \input{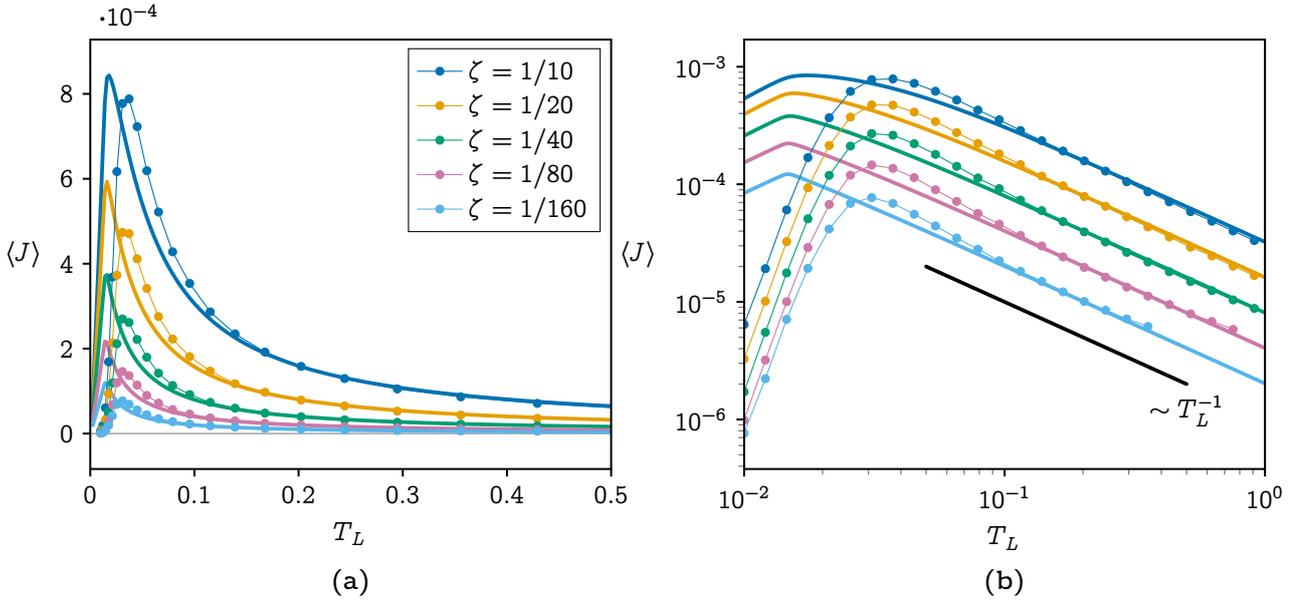}%

    \caption{Simulation result for the mean current through the chain for several values of the friction coefficient of the bath particles $\zeta$. Lower friction implies better time-scale separation. \textbf{(a)} linear scale, with comparison to theoretical prediction (solid line). \textbf{(b)} log-log plot which displays $\langle J\rangle \propto T_L^{-1}$ behavior at high $T_L$.}
    \label{fig:nondim_comparison}
\end{figure}
\begin{figure}[!htbp]
    \centering
    \tikzsetnextfilename{dimfulT2}%
    \input{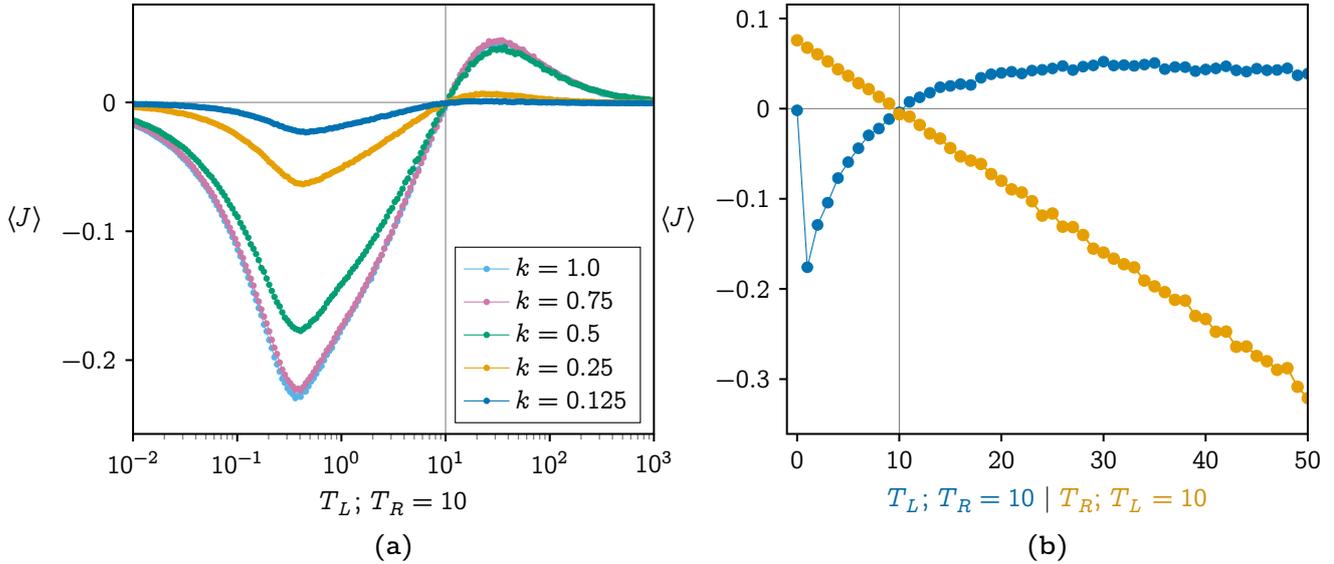}%

    \caption{Simulation results of the mean current through the chain, now with $T_L\neq T_R$ to disentangle $T_{L}$ and $\Delta=\abs{T_{R}-T_{L}}$. \textbf{(a)} Mean current for the chain for various spring constants, and $T_R=10$. Nonmonotonic behavior is also observed when $T_L<T_R$, \textbf{(b)} Juxtaposition of the respective currents when $T_L$ and $T_R$ are kept constant. Varying $T_L$ produces nonlinear dependency of the current on the temperature; varying $T_R$ reproduces the typical linear behavior.}
    \label{fig:vary_T2}
\end{figure}

Extending the obtained effective description of the left most oscillator, we rewrite the dynamics of the full chain Eq.~\eqref{eq:EOMq1} in the regime of time-scale separation and weak coupling as,
\begin{subequations}\label{eq:effecftivechain}
\begin{align}
m\ddot q_1 &= -k_\text{eff} q_1 - k(q_1 - q_2) - \gamma_L \dot q_1 + \sqrt{2\gamma_LT_L}\eta_L, \\
m\ddot q_i &= -k(2q_i - q_{i-1} - q_{i+1}), \hspace{4cm} i = 2, \dots, N-1, \\
m\ddot q_N &= -K q_N - k(q_N - q_{N-1}) - \gamma_R \dot q_N + \sqrt{2\gamma_LT_L}\eta_R(t).\label{eqm}
\end{align}
\end{subequations}
where $\eta_R$ and $\eta_L$ are independent standard white noises.
The renormalized stiffness 
\begin{equation}
    k_\text{eff}=K - N_b\,\varepsilon^2 \frac{\lambda^2}{T_L}
\left(\frac{I_0(2p)}{I_0(p)^2} - 1\right),
\end{equation}
and friction coefficient 
\begin{equation}
    \gamma_L =
N_b\varepsilon^2 \frac{\zeta L^2}{2\pi^2 T_L^2}
\left(\frac{\lambda}{I_0(p)}\right)^2
\sum_{n=1}^{\infty} \frac{I_n(p)^2}{n^2}
\end{equation} 
follow from the results of the previous previous section, $\eta_{L,R}$ are mutually independent white noises.
The current can be computed using the nonequilibrium Greens functional formalism~\cite{dhar2014heat}, 
The stationary current, derived in Appendix~\ref{app:current}, is
\begin{equation}\label{eq:J_closed_form}
J= \frac{4\gamma_L\gamma_R k^2}
{m\bigl(D_0-D_1+\sqrt{D_0(D_0-2D_1)}\bigr)}\,(T_L-T_R)
\end{equation}
where
\begin{align}
D_0 = \gamma_LK^2+\gamma_Rk_\text{eff}^2,\quad D_1 = 2k\Big[\gamma_L(K-k)+\gamma_R(k_\text{eff}-k)\Big]
-\frac{2k}{m}\gamma_L\gamma_R(\gamma_L+\gamma_R).\label{eq:d0d1}
\end{align}
\textcolor{black}{Therefore}, the stationary current 
can be written as $J=J_0(T_L)\,(T_L-T_R)\,$, where the prefactor $J_0$ depends on the temperature $T_L$ of the particle reservoir through the effective coefficients $k_\text{eff}$ and $\gamma_L$. 
Using
$\gamma_L\sim T_L^{-2}$ and $k_\text{eff}=K+\mathcal O(T_L^{-1})$, we find the asymptotic behavior $ J_0(T_L)\sim T_L^{-2}$ as $T_L\to \infty$.  
It implies that for a fixed $T_R< T_L$, the stationary current $J\sim T_L^{-1}$, as $T_L\to\infty$, implying that the current is non-monotonic in  $T_L$ vanishing in both the limits $T_L\to T_R$ and $T_L\to\infty$, attaining a maximum in between. 
Beyond the maximum, the current decreases on increasing the drive, leading to a negative differential conductivity (NDC). The emergence of the NDC in a linear system
without any kinetic constraints or active elements~\cite{maes1996effect,zia2002getting,PhysRevE.84.061108,baerts2013frenetic,li2006negative,leitmann2013nonlinear,santra2022activity} in the bath sets it apart from the few similar phenomena observed previously.

These conclusions are reinforced by simulation results (see Fig.~\ref{fig:nondim_comparison}), which show a good qualitative match with Eq.~\ref{eq:J_closed_form} and a clear $\langle J\rangle\propto T_L^{-1}$ regime at high $T_L$.
In Fig.~\ref{fig:vary_T2}a, $T_R\neq0$ and the case $T_L<T_R$ is showcased.
In addition, in Fig.~\ref{fig:vary_T2}b, it is confirmed that varying $T_R$ recovers the typical linear behavior. \textcolor{black}{It must also mentioned that though the analytical calculations were done in the fast bath, and weak coupling limit, the negative differential conductivity exists for finite couplings as well.}

\section{Conclusion}
Oscillator chains have been central in physics for the study of energy transport and diffusion.  
Historical highlights include the foundational work on the Fermi-Pasta-Ulam-Tsingou problem \cite{berman2005fpu}, and Peierls’ theory of heat conduction in crystals \cite{peierls1929kinetictheory}. 
\emph{Oscillators} have typically referred to vibrational degrees of freedom (phonons) such as found in crystals and even more generally, albeit strongly attenuated, in visco-elastic materials.

The present study has kept the conducting medium as a harmonic chain but replaces one of the reservoirs by a particle bath with distinct, purely positional degrees of freedom. 
That change in the nature of the bath and in the way it couples to the system turns out to be sufficient to produce the observed transport behavior, characterized by negative differential conductivity: even when the particle bath is thermal and un-driven and the medium is linear, we find that the resulting heat current differs markedly from that in purely oscillator-based setups. 
The origin is thermokinetic and  highlights the importance of modeling the environment and its coupling carefully: open far-from-equilibrium transport is not determined by the conducting medium alone, but by the interplay between the system and the reservoirs to which it is connected.

\textcolor{black}{There are several natural directions for future work. It is interesting to replace the passive particle bath by an active one, for example a bath of run-and-tumble particles or active Brownian particles, and investigate how activity modifies the effective friction and the resulting heat transport. The thermokinetic effect identified here is also expected to play a central role in nonequilibrium response, in particular in the excess heat associated with quasistatic changes of the bath temperature, which determines the corresponding nonequilibrium heat capacity. Finally, the soft-matter baths introduced in this work provide a minimal framework that could be extended to model active or fluctuating membranes.
}

\vspace{1cm}
\noindent 
{\bf Acknowledgment:}
IS acknowledges funding from the European Union's
Horizon 2024 research and innovation programme
under the Marie Sklodowska--Curie (HORIZONTMAMSCA-PF-EF) grant agreement No. 101205210.
\newpage

\appendix
\renewcommand{\theequation}{\thesection \arabic{equation}}
\setcounter{equation}{0}
\begin{center}
\Large \textbf{Appendix}
\end{center}
\section{Effective bath}\label{fric}
For a single probe in the particle bath, the governing equations are
\begin{align}
\zeta \dot x_i(t)
&=
-\varepsilon  q(t)\, \mathcal F'(x_i(t))
+ \sqrt{2\zeta T_L}\,\xi_i(t),
\qquad i=1,\dots,N_b,
\\
m \ddot q(t)
&=
- K q(t)
- \varepsilon \sum_{i=1}^{N_b} \mathcal F(x_i(t)).
\end{align}
We decompose the force from the $i$th bath particle acting on the probe into three parts,
\begin{align}
F^{i}_{\mathrm{bath}}(t)
=
F_i^{\mathrm{qs}}(t)
+
F_i^{\mathrm{fric}}(t)
+
\eta_i(t),
\end{align}
where
\begin{align}
F_i^{\mathrm{qs}}(t)
&=
-\varepsilon\la \mathcal F(x_i)\ra_{q(t)},
\\
F_i^{\mathrm{fric}}(t)
&=
-\varepsilon\Big(
\la \mathcal F(x_i(t)) \mid q(s),\, s\le t\ra
-
\la \mathcal F(x_i)\ra_{q(t)}
\Big),
\\
\eta_i(t)
&=
-\varepsilon\Big(
\mathcal F(x_i(t))
-
\la \mathcal F(x_i(t)) \mid q(s),\, s\le t\ra
\Big).
\end{align}
The first term is the mean force acting on the probe in the quasistatic limit.
The average $\la\cdots\ra_{q(t)}$ denotes averaging with the probe held fixed at position $q(t)$.
The average $\la \cdots \mid q(s),\, s\le t\ra$ denotes averaging over bath trajectories evolving under the history of the probe trajectory up to time $t$.
The second term can be interpreted as the response in the mean force experienced by the probe due to its own past motion, and the remaining third term is the noise.
We now proceed to treat these terms separately.

\subsection{Mean force}
For a frozen probe position, the bath distribution is given by
\begin{align}
\rho_q(x)
&=
\mathcal Z^{-1}
e^{-\frac{\varepsilon q\, \mathcal F(x)}{T_L}},
\qquad
\text{where}
\qquad
\mathcal Z
=
\int_0^L \dd{x}\,
e^{-\frac{\varepsilon q\, \mathcal F(x)}{T_L}}
\\
&=
\frac{1}{L}\left[
1-\frac{\varepsilon q}{T_L}
\big(\mathcal F(x)-\la \mathcal F\ra_0\big)
+O(\varepsilon^2)
\right].
\label{eq:rhoeps}
\end{align}
where $\la\cdots\ra_0$ denotes averages with respect to $\rho_q(x)$ at $\varepsilon=0$.
Using this,
\begin{align}
F_i^{\mathrm{qs}}(t)
&=
-\varepsilon \la \mathcal F\ra_0
+ \varepsilon^2\,\frac{ q(t)}{T_L}\,
\Big(\la \mathcal F^2\ra_0-\la \mathcal F\ra_0^2\Big)
+O(\varepsilon^3)
\\
&=
-\varepsilon \lambda
+\varepsilon^2\lambda^2\,\frac{ q(t)}{T_L }\,
\left(
\frac{I_0(2p)}{I_0(p)^2}-1
\right)
+O(\varepsilon^3).
\end{align}
The term inside the brackets in the second term is always positive, implying that the interactions with the bath decrease the effective pinning of the probe.

\subsection{Friction}
To evaluate the friction, we need the response of $\mathcal F(x)$ to the perturbation
$\varepsilon [q(s)-q(t)]\,\mathcal F(x)$.
Using the Kubo formula, we get
\begin{align}
F_i^{\mathrm{fric}}(t)
&=
\frac{\varepsilon}{T_L}
\int_{\mathrlap{-\infty}}^t \dd{s}\,
\left\la
\mathcal F(x_t);
\partial_s\Big(-\varepsilon [q(s)-q(t)]\mathcal F(x_s)\Big)
\right\ra
\\
&=
-\frac{\varepsilon^2}{T_L}
\int_{\mathrlap{-\infty}}^t \dd{s}\,
\dot q(s)\,
\left\la \mathcal F(x_t);\mathcal F(x_s)\right\ra_q .
\end{align}
Using Eq.~\eqref{eq:rhoeps}, the above reduces to
\begin{align}
F_i^{\mathrm{fric}}(t)
&=
-\frac{\varepsilon^2}{T_L}
\int_{\mathrlap{-\infty}}^t \dd{s}\,
\dot q(s)\,
\left\la \mathcal F(x_t);\mathcal F(x_s)\right\ra_0
+O(\varepsilon^3).
\end{align}
So we need to evaluate the above correlation for the stochastic process $x$, which is a free diffusion on a ring of length $L$.
It is useful to rewrite $\mathcal F(x)$ using the Jacobi--Anger expansion
\begin{align}
\mathcal F(x)
=
\frac{\lambda}{I_0(p)}
\left(
I_0(p)+2\sum_{n=1}^\infty I_n(p)\cos(2\pi n x/L)
\right)
\end{align}
and use the identity~\cite{santra2021active}
\begin{align}
\la \cos(2\pi m x_t/L)\cos(2\pi n x_s/L)\ra_0
=
\frac12\delta_{m,n}
e^{-\frac{4\pi^2n^2T_L}{\zeta L^2}|t-s|}
\end{align}
to obtain
\begin{align}
\la \mathcal F(x_t);\mathcal F(x_s)\ra_0
=
\frac{2\lambda^{2}}{I_0(p)^2}
\sum_{n=1}^{\infty}
I_n(p)^2
e^{-\frac{4\pi^2n^2T_L}{\zeta L^2}|t-s|}.
\end{align}
For large $T_L$, the exponential decays very fast compared to the slow probe, and we obtain a Markovian friction
\begin{align}
\gamma_i^{\mathrm{eff}}
=
\varepsilon^2\frac{\zeta L^2}{2\pi^2T_L^2}
\left(\frac{\lambda}{I_0(p)}\right)^2
\sum_{n=1}^\infty\frac{I_n(p)^2}{n^2}.
\end{align}

\subsection{Noise}
Because the system with a single probe is in equilibrium, the noise correlations follow a second fluctuation--dissipation relation,
\begin{align}
\la \eta_i(t)\eta_j(s)\ra
=
2\gamma_i^{\mathrm{eff}}T_L\,\delta_{ij}\delta(t-s).
\end{align}

\section{Current computation}\label{app:current}
It is more convenient for the computations to rewrite \eqref{eq:effecftivechain},
\begin{equation}
M \ddot X(t) = -\Phi X(t) - \Gamma \dot X(t) + \Xi(t),
\end{equation}
where 
\begin{align}
M = m \mathbb{I},\qquad \Gamma = \Gamma_L + \Gamma_R,\quad \text{ with  }
(\Gamma_L)_{ij} = \gamma_L\delta_{i1}\delta_{j1}, \qquad
 (\Gamma_R)_{ij} = \gamma_R\delta_{iN}\delta_{jN},
\end{align}
and $\Phi$ is the tridiagonal force matrix with entries
\begin{align}
\Phi_{ij}
= \left(2k + (k_{\text{eff}} - k)\,\delta_{i1} + ( K - k)\,\delta_{iN}\right)\delta_{ij}
\;-\; k\left(\delta_{i,j+1} + \delta_{i,j-1}\right),
\end{align}
and $\Xi(t) = (\xi_L(t), 0, \dots, 0, \xi_R(t))^T$.

The general solution in Fourier space is given in terms of the Green's function $G(\omega)$,
\begin{equation}
\widetilde X(\omega) = G(\omega)\,\widetilde\Xi(\omega),\quad\text{with }G(\omega) = \big[-M\omega^2 + \Phi - i\omega \Gamma \big]^{-1}.
\label{eq:solx}
\end{equation}

The heat current injected by the particle bath is
\begin{equation}
J = \la \dot q_1(-\gamma_L\dot q_1(t)+\xi_L(t)-k_{\text{eff}}q_1(t))\ra= \big\langle \dot X^T(t)\big[-\Gamma_L \dot X(t) + \Xi_L(t)\big] \big\rangle,
\end{equation}
where $\Xi_L(t) = (\xi_L(t), 0, \dots, 0)^T$, and in writing the second equality we have used $\la q_1\dot q_1\ra=0$ in the stationary state (this can be shown that the resulting Fourier space integrand for this is an odd function in $\omega$, and hence zero).

The steps A12-A17 of Ref.~\cite{santra2022activity} can be carried out exactly in the same way, and leads to,
\begin{equation}
J = 2(T_L - T_R)\int_{-\infty}^{\infty} \frac{\dd{\omega}}{2\pi}\,
\omega^2\, \mathrm{Tr}\!\left[G(\omega)\Gamma_R G^*(\omega)\Gamma_L\right].
\end{equation}
Using the explicit forms of $\Gamma_{R,L}$, this reduces to,
\begin{equation}
J =
2\gamma_L \gamma_R (T_L - T_R)
\int_{-\infty}^{\infty} \frac{\dd{\omega}}{2\pi}\,
\omega^2\, |G_{1N}(\omega)|^2.
\end{equation}
So to obtain the stationary current, we need to evaluate the Greens function matrix element $G_{1N}$.

Noting that $G^{-1}(\omega)$ is a tridiagonal matrix, with off diagonal entries $-k$, and
\begin{align}
a_1 &= k+k_{\text{eff}}-m\omega^2-i\omega\gamma_L,\\
a_j =a  &= 2k-m\omega^2,\quad j=1,\cdots N-1\\
a_N &= k+ K-m\omega^2-i\omega\gamma_R.
\end{align}
For a tridiagonal matrix~\cite{usmani1994inversion},
\begin{equation}
    G_{lN}=(-k)^{N-l}\frac{\theta_{l-1}}{\theta_N},
\end{equation}
where 
\begin{align}
\theta_l &= a\,\theta_{l-1}-k^2\theta_{l-2},
\qquad l=2,3,\dots,N-1,\\
\theta_N &= a_N\,\theta_{N-1}-k^2\theta_{N-2}.
\end{align}
Using the boundary conditions $\theta_0 = 1,
\theta_1 = a_1$, our required element is,
\begin{equation}
G_{1N}(\omega)=\frac{(-k)^{N-1}}{\theta_N}.
\end{equation}
The bulk recursion has the solution,
\begin{equation}
\theta_l=
\frac{(-k)^{\,l-1}}{\sin q}
\Big[
k\sin\big((l+1)q\big)+\Delta_L\sin(lq)
\Big],
\qquad l=1,2,\dots,N-1.
\label{eq:bulktheta}
\end{equation}
where 
\begin{equation}
\Delta_L \equiv a_1-a=k_{\text{eff}}-k-i\omega\gamma_L.
\end{equation}
Using this in $\theta_N$ expression,
\begin{equation}
\theta_N=
\frac{(-k)^N}{\sin q}
\Big[
\mathcal A(q)\sin(Nq)+\mathcal B(q)\cos(Nq)
\Big]
\end{equation}
where $\Delta_R = K-k-i\omega\gamma_R$, $\omega=\omega_c\sin\frac{q}{2}$ with $
\omega_c=2\sqrt{\frac{k}{m}}$ and
\begin{align}
\mathcal A(q)
&=
\frac{\Delta_L+\Delta_R}{k}
+
\left(1+\frac{\Delta_L\Delta_R}{k^2}\right)\cos q,\\
\mathcal B(q)
&=
\left(1-\frac{\Delta_L\Delta_R}{k^2}\right)\sin q.
\end{align}
Thus we have,
\begin{equation}
G_{1N}(\omega)
=
-\frac{\sin q}{k\big[\mathcal A(q)\sin(Nq)+\mathcal B(q)\cos(Nq)\big]}.
\end{equation}

We now consider the limit of a thermodynamically large chain $N\to\infty$.
For $\omega>\omega_c$, $q$ becomes complex, and the contribution is exponentially small in $N$.
Thus, in the thermodynamic limit, we may restrict to $0\le \omega\le \omega_c$,
equivalently $0\le q\le \pi$, and average over the fast phase $x=Nq$, which leads to, 
\begin{align}
J &\simeq
\frac{4\gamma_L\gamma_R(T_L-T_R)}{\pi k^2}
\int_0^{\omega_c} \dd{\omega}\omega^2\sin^2 q
\int_0^{2\pi}\frac{dx}{2\pi}\,
\frac{1}{|\mathcal A(q)\sin x+\mathcal B(q)\cos x|^2}.\\
&=\frac{4\gamma_L\gamma_R(T_L-T_R)}{\pi k^2}
\int_0^{\omega_c} \dd{\omega}
\frac{\omega^2\sin^2 q}{\big|\mathrm{Im}(\mathcal A(q)\mathcal B^*(q))\big|}
\end{align}
where we use the identity 
\begin{equation}
\int_0^{2\pi}\frac{\dd{x}}{2\pi}\,
\frac{1}{\alpha \sin^2 x+\beta \cos^2 x+2c\sin x\cos x}
=
\frac{1}{\sqrt{\alpha\beta-c^2}}
\end{equation}
with $\alpha=|\mathcal A|^2,\quad
\beta=|\mathcal B|^2,\quad
c=\mathrm{Re}\big[\mathcal A\mathcal B^*\big]$.

The denominator can be further simplified,
\begin{equation}
\mathrm{Im}\!\left[\mathcal A(q)\mathcal B^*(q)\right]
=
-\frac{\omega\sin q}{k^3}\left(\gamma_L\Big[ K^2-m\omega^2( K-k)\Big]
+\gamma_R\Big[k_{\text{eff}}^2-m\omega^2(k_{\text{eff}}-k)\Big]
+\omega^2\gamma_L\gamma_R(\gamma_L+\gamma_R).\right),
\end{equation}
Thus the final expression in the integral form, using $\left|\frac{d\omega}{dq}\right|
=
\frac{k\sin q}{m\omega}$, is
\begin{equation}
J =
\frac{4\gamma_L\gamma_R k^2 (T_L-T_R)}{\pi m}
\int_0^\pi \dd{q}
\frac{\sin^2 q}{\left(\gamma_L Z_R(\omega)
+\gamma_RZ_L
+\omega^2\gamma_L\gamma_R(\gamma_L+\gamma_R).\right)}.
\end{equation}
with 
\begin{align}
    Z_{R,L}(\omega)&=\Big[\lambda_{R,L}^2-m\omega^2(\lambda_{R,L}-k)\Big]
    \label{eq:capz}
\end{align}
Finally, performing the integral yields the closed-form expression found in the main text
\begin{equation}
J= \frac{4\gamma_L\gamma_R k^2(T_L-T_R)}
{m\bigl(D_0-D_1+\sqrt{D_0(D_0-2D_1)}\bigr)},
\end{equation}
where $D_0,\,D_1$ are defined in the main text.

\printbibliography
\end{document}